\documentclass[%
reprint,
superscriptaddress,
showpacs,%
showkeys,
nofootinbib,
nobibnotes,
amsmath,amssymb,
aps,
pre,
longbibliography,
floatfix,
]{revtex4-1}

\usepackage[utf8]{inputenc}
\usepackage[T1]{fontenc}
\usepackage[english]{babel}
\usepackage{amsfonts}
\usepackage{amsmath}
\usepackage{graphicx}
\usepackage{color}
\usepackage{hyperref}

\hypersetup{colorlinks = true, allcolors = blue}

\newcommand{\widfig}[2]{\includegraphics[width=#1\columnwidth]{#2.eps}}

\newcommand{\myii}{\mathrm{i}}
\newcommand{\myexp}{\mathrm{e}}
\newcommand{\mydd}{\mathrm{d}}

\newcommand{\mean}[1]{\left\langle #1\right\rangle}

\newcommand{\tilkappa}{\tilde{\kappa}}
\newcommand{\inxcut}{\text{cut}}
\newcommand{\argA}{\Phi}

\begin{document}

\title{Hyperbolic chaos at blinking coupling of noisy oscillators}

\author{Pavel V. Kuptsov}\email[Electronic
address:]{p.kuptsov@rambler.ru}%
\affiliation{Department of Instrumentation Engineering, Saratov State
  Technical University,\\ Politekhnicheskaya 77, Saratov 410054,
  Russia}%
\affiliation{Institute of Physics and Astronomy, University of
  Potsdam, Karl-Liebknecht-Str. 24/25, 14476 Potsdam-Golm, Germany}%
\author{Sergey P. Kuznetsov}%
\affiliation{Kotel’nikov’s Institute of Radio-Engineering and
  Electronics of RAS, Saratov Branch,\\ Zelenaya 38, Saratov 410019,
  Russia}%
\affiliation{Institute of Physics and Astronomy, University of
  Potsdam, Karl-Liebknecht-Str. 24/25, 14476 Potsdam-Golm, Germany}%
\author{Arkady Pikovsky}%
\affiliation{Institute of Physics and Astronomy, University of
  Potsdam, Karl-Liebknecht-Str. 24/25, 14476 Potsdam-Golm, Germany}%



\date{\today}

\begin{abstract}
  We study an ensemble of identical noisy phase oscillators with a
  blinking mean-field coupling, where one-cluster and two-cluster
  synchronous states alternate. In the thermodynamic limit the
  population is described by a nonlinear Fokker-Planck equation. We
  show that the dynamics of the order parameters demonstrates
  hyperbolic chaos. The chaoticity manifests itself in phases of the
  complex mean field, which obey a strongly chaotic Bernoulli
  map. Hyperbolicity is confirmed by numerical tests based on the
  calculations of relevant invariant Lyapunov vectors and Lyapunov
  exponents. We show how the chaotic dynamics of the phases is
  slightly smeared by finite-size fluctuations.
\end{abstract}

\maketitle

\section{Introduction}

Dynamics of large
populations of coupled oscillators attracted large interest recently.
Problems of this type appear in studies of Josephson junctions, lasers,
ensembles of neurons, cell populations, and many other fields. From
a more general perspective, studies of such a system allows one to shed light on
a long-standing problem of an interrelation between microscopic and macroscopic
dynamics. Indeed, the mostly studied nontrivial effect in the ensembles
of globally coupled oscillators
is their synchronization, which can be considered as a nonequilibrium phase
transition~\cite{Kuramoto-84,Pikovsky-Rosenblum-Kurths-01}. Remarkably,
in some situations one can explicitly derive the dynamics of global variables
(order parameters), in terms of which the synchronization transition is a bifurcation
from zero equilibrium to a non-trivial state~\cite{Ott-Antonsen-08}.

Different interrelations between regularity properties on micro- and
macro-levels (i.e. for individual oscillators and for mean fields)
have been reported in the literature. For example, chaotic
micro-oscillators being coupled may lead to periodic mean
fields~\cite{Kaneko-92b,Pikovsky-Kurths-94a,Pikovsky-Rosenblum-Kurths-96,
  Topaj-Kye-Pikovsky-01,Ott-So-Barreto-Antonsen-02}. On the other
hand, coupled periodic oscillators may produce chaotic mean
fields~\cite{Hakim-Rappel-92,Nakagawa-Kuramoto-94, KuzPikRos10}.
While description of populations of deterministic oscillators remains
a challenging task, there exists a nice framework for ensembles of
noise-driven oscillators. The behavior in the thermodynamic limit can
be described self-consistently with a nonlinear Fokker-Planck
equation, first suggested by Desai and Zwanzig~\cite{Desai-Zwanzig-78}
in the context of globally coupled noisy bistable oscillators (see
also~\cite{Dawson-83}). This approach has been then applied to noisy
periodic oscillators in
Refs.~\cite{Bonilla-Neu-Spigler-92,Pikovsky-Ruffo-99}.

In this paper we consider a population of noisy oscillators subject to blinking, time-periodic
coupling~\cite{Belykh_etal-04,Wang-Shi-Sun-10,Porfiri-12}. This is a minimal generalization of the
simplest model with constant coupling, which demonstrates simple synchronization patterns only.
We will show that with a blinking coupling, where on different periods of the total cycle different
synchronous modes emerge, the total dynamics demonstrates highly nontrivial regime of phase hyperbolic
chaos. In this regime the phase of the complex order parameter obeys a doubling Bernoulli map, which has
strong chaotic properties, and is, contrary to many other models of chaos, structurally stable in respect to
perturbations.

Our treatment of the dynamics of the complex mean fields follows
recent studies~\cite{KuzBook,KuzUfn}, where a general framework for mechanisms
of hyperbolic chaos in coupled oscillators has been developed.
In the simplest setup, two oscillators that are excited alternately
can interact in a way to influence the phases of each others at the stages
where these oscillators pass through a Hopf bifurcation. More precisely,
one needs that the appearing field is forced by a higher harmonic of
another oscillator, then a transformation of the phase $\phi\to n\phi$
with $|n|\geq 2$ can occur. As the result the phase obeys a
Bernoulli-type, uniformly expanding map, and the whole strange attractor is of
famous Smale-Williams type~\cite{Smale67}.
The described mechanism of phase multiplication generating hyperbolic
chaos is quite generic. The straightforward way of its implementation
is to consider two non-autonomous self oscillators with different
natural frequencies and appropriate coupling~\cite{Hyp}. Further
analysis revealed a possibility to identify this mechanism in a
system of two or three coupled autonomous
oscillators~\cite{KuznetsovPikovsky}. Recently it was described,
 how to
observe the hyperbolic chaos in a spatially extended system, as
a result of
interaction of the Turing modes~\cite{HypTur2012}.
In paper~\cite{KuzPikRos10} an alternation between
synchronized
and desynchronized regimes in two ensembles of non-identical oscillators
subjected to the Kuramoto
transition was shown to possess a collective hyperbolic
chaos of complex order parameters.

In this paper we study an
ensemble of identical stochastic phase oscillators coupled via the mean
fields. We consider a situation of blinking coupling, where
different synchronization patterns,
one with one cluster and another with two clusters, alternate.
We demonstrate,
both basing on the nonlinear Fokker-Planck equation, and on the direct
simulation of a large population that the phases of these patterns
demonstrate hyperbolic chaos. Moreover, we study the Lyapunov exponents
of the system and apply numerical criteria based on Lyapunov
vectors to verify hyperbolicity.

\section{Ensemble of coupled noisy oscillators}
\subsection{Microscopic equations}

Our basic model is an ensemble of $K$  globally coupled
identical limit-cycle oscillators with
additive noise. In the phase
description the ensemble is governed by the following set of
equations:
\begin{equation}
    \label{eq:ens}
    \dot\phi_k=f(\phi_k)
    +\frac{1}{K}\sum_{j=1}^K q(\phi_j-\phi_k)+\sigma\xi_k(t),
\end{equation}
where $k=1,\ldots,K$, $\phi_k$ is the phase of a $k$-th oscillator (taken in the reference frame rotating with the
basic frequency),  and the independent random variables $\xi_k(t)$ describe a
white Gaussian noise:
$\mean{\xi_k(t)\xi_j(t+\tau)}=\delta_{jk}\delta(\tau)$.
Functions $q(\phi)$ and $f(\phi)$ (both are $2\pi$-periodic) describe, respectively, effects of global coupling and of
the external periodic forcing at the oscillator frequency and/or its harmonics.
It is convenient to define these functions by means of the Fourier decompositions
\begin{equation}
  \label{eq:four_fun}
    f(\phi)=\sum_{n=-\infty}^{\infty}F_n\myexp^{\myii n\phi}, \;\;
    q(\phi)=\sum_{n=-\infty}^{\infty}Q_n\myexp^{\myii n\phi}.
\end{equation}
Since the functions $f$ and $q$ are real, $F_{-n}=F_n^*$, and
$Q_{-n}=Q_n^*$, where asterisk denotes complex conjugation.

Given a state of the ensemble, one can determine the following mean fields
playing roles of order parameters:
\begin{equation}
  \label{eq:order_prm}
  A_\ell(t)=\langle\myexp^{\myii \ell\phi} \rangle =\frac{1}{K}\sum_{k=1}^K \myexp^{\myii \ell\phi_k},
\end{equation}
where $\ell=1,2,\ldots$, and $A_0=1$.  Notice that
$A_{-\ell}=A_\ell^*$. In what follows we shall denote $\arg
A_\ell=\argA_\ell$.

\subsection{Nonlinear Fokker-Plank equation}

In the thermodynamic limit $K\to\infty$ the ensemble can be described
by the density of the probability distribution $v(\phi,t)$. The mean
fields in this case are expressed as
\begin{equation}
  \label{eq:order_prm_thrm_lim}
  A_\ell(t)=\int_0^{2\pi}v(\phi,t)\myexp^{\myii \ell\phi}\mydd\phi.
\end{equation}
The probability density can be decomposed as Fourier series:
\begin{equation}
  \label{eq:prob_dens}
  v(\phi,t)=\frac{1}{2\pi}
  \sum_{\ell=-\infty}^\infty A_\ell(t)\myexp^{-\myii \ell\phi},
\end{equation}
i.e., the order parameters $A_\ell(t)$ are just coefficients of the
Fourier modes for the probability density. 

The dynamics of the probability density can be described by the
nonlinear Fokker-Planck
equation~\cite{Desai-Zwanzig-78,Dawson-83}. For the
system~\eqref{eq:ens}, accounting the expressions~\eqref{eq:four_fun},
we can write out this equation as
\begin{equation}
  \label{eq:fp}
  \frac{\partial v}{\partial t}=-\frac{\partial}{\partial \phi}
  \sum_{n=-\infty}^{\infty}v(F_{-n}+Q_nA_n)\,\myexp^{-\myii n\phi}
  +\frac{\sigma^2}{2}\frac{\partial^2v}{\partial\phi^2}.
\end{equation}
Now we substitute the Fourier decomposition~\eqref{eq:prob_dens} for
$v(\phi,t)$, then multiply the equation by $\myexp^{-\myii \ell\phi}$
and integrate it over $2\pi$. The resulting equations for $A_\ell$
read:
\begin{equation}
  \label{eq:modes_comm}
    \dot A_\ell=\myii\ell \sum_{n=-\infty}^{\infty}
    (F_{-n}+Q_nA_n)A_{\ell-n} -\frac{\sigma^2\ell^2}{2}A_\ell.
\end{equation}

\subsection{Elementary synchronization dynamics}
\label{sec:esd}

The set of equations \eqref{eq:modes_comm} allows one a simple
qualitative description of basic synchronization phenomena in the
ensemble. If only the global coupling is present ($F_n=0$), and
$q(\phi)$ has one Fourier component $Q_n=-\frac{\myii}{2}\kappa_n$,
then the non-synchronized state $A_j=\delta_{j,0}$ remains stable
until the coupling passes the synchronization threshold, 
\begin{equation}
  \label{eq:cluster_gr_cond}
  \kappa_n>n\sigma^2.
\end{equation}
Above this threshold, the mode $A_n$ becomes unstable, and a
stationary solution of \eqref{eq:modes_comm} containing modes $A_{\ell
  n}$ with $\ell=1,2,3,\ldots$ establishes.  This corresponds to the
appearance of an $n$-clustered state of the ensemble, where the
probability density has maxima at $\phi^{(0)}+k2\pi/n$, and
$\phi^{(0)}$ is an arbitrary phase shift determined by initial
conditions. In contrast, without the global coupling ($Q_n=0$), if
only the external forcing is present containing one Fourier component
($F_n=-\frac{\myii}{2}\epsilon$), then for small $\epsilon$ an
$n$-cluster state with
$A_n\approx -\epsilon n^{-1}\sigma^{-2}$ 
appears. The phase of this
cluster is fixed, been determined by the applied forcing.

\section{Cluster exchange dynamics at blinking coupling}
The main goal of this paper is to describe macroscopic chaos appearing
when the coupling in the ensemble is blinking
(cf.~\cite{Belykh_etal-04}).

In this section we consider the effect of blinking coupling
qualitatively, basing on the elementary synchronization dynamics
described in the previous section. First, we make a particular choice
for the global coupling terms, assuming alternating couplings $Q_1$
and $Q_2$:
\begin{equation}
  \label{eq:coeffs}
  Q_{1,2}=-\frac{\myii\tilkappa_{1,2}(t)}{2},
\end{equation}
where
\begin{equation}
  (\tilkappa_{1},\tilkappa_{2})=
  \begin{cases}
    (\kappa_1,0), & \text{if $mT\leq t<mT+T/2$},\\
    (0,\kappa_2), & \text{if $mT+T/2\leq t<mT+T$}.
  \end{cases}
  \label{eq:blcoupl}
\end{equation}
Here $T$ is the total period of blinking; it is assumed to be large
enough compared to characteristic time of cluster formation or
decay. In the first half of this period (we call it stage 1) only
$Q_1$ is present, correspondingly a one-cluster state develops; during
the second half of the period (stage 2) only $Q_2$ is present, so in
this stage the two-cluster state develops (see Fig.~\ref{fig:sptm_fp_phs}
below).

For the following consideration, the phases of the clusters are important. As the one-cluster state contains
all modes $A_n$, $n=1,2,3,\ldots$, at the beginning of the stage 2 the amplitude $A_2\sim A_1^2$ is finite.
On the stage 2, due to the
appropriate coupling the instability appears associated with formation of the two-cluster state, and the initial phase
$\argA_2=2\argA_1$ occurs to be definite, determined by the preceding evolution. Thus,
the two-cluster state accepts this phase from the former  one-cluster state.
At the end of stage 2, only modes with even index $A_n$, $n=2,4,6,\ldots$ are present (as $T$ is large enough, the odd modes decay effectively during the stage 2). Thus, at the beginning of the next stage 1 there is no initial amplitude $A_1$, and its growth due to the coupling $Q_1$ would start only from random initial fluctuations, and the appearing one-cluster state would have a random phase.

The situation changes if a small regular external forcing is present. As one can see
from  \eqref{eq:modes_comm}, the forcing $F_n$ provides appearance of terms $\sim iF_{-n}A_{\ell -n}$, thus
producing combinational modes.

Let us consider the effect of the mode $F_3=-\frac{\myii\epsilon}{2}$. At the
stage 1 it produces
combinational modes with all indices, but because these modes are already
present in the distribution, the
effect of small $\epsilon$ is negligible. At the stage 2, however, the mode $A_1$ is produced by terms
$F_3^*A_2^*+F_3A_4$ in  \eqref{eq:modes_comm}. Because $A_2$ is typically much larger than
$A_4$, the result of this interaction is the appearance of a small but notable amount of mode 1
with $A_1\sim \epsilon A_2^*$ at the end of the stage 2.

This circumstance changes essentially the starting conditions for the evolution on the stage 1 compared to the case $\epsilon=0$:
now the growing mode $A_1$ develops from the seed
$\sim \epsilon A_2^*$, and so its phase will be $\argA_1=-\argA_2$.
Now the phases of cluster states arising at all stages of evolution are non-random, but depend on the previous phases in a deterministic way. Combining the transformations $\argA_2=2\argA_1$ from stage 1 to stage 2, and $\argA_1=-\argA_2$ from stage 2 to stage 1, we obtain the Bernoulli-type map
\begin{equation}
  \label{eq:ber}
  \argA_1(m+1)=-2\argA_1(m)
\end{equation}
at successive periods of the blinking coupling.

The map \eqref{eq:ber} for the phase transformation is uniformly expanding and hyperbolic,
and we conclude from these qualitative arguments that the cluster patterns for blinking coupling of
noisy oscillators in the situation we have considered will demonstrate hyperbolic chaos in the dynamics of the order parameters (the modes of the distribution density).

\section{Illustrations and characterization of the dynamics}

In this section we confirm numerically the existence of the hyperbolic chaos
in the setup described above. The mode equations take
the form:
\begin{equation}
  \label{eq:model_fp}
  \begin{split}
    \dot A_\ell=&\frac{\ell}{2}\Biggl[\sum_{n=1}^2
    \tilkappa_n\left(A_nA_{\ell-n}-A_{-n}A_{\ell+n}\right)\\
    {}&+\epsilon(A_{\ell+3}-A_{\ell-3})\Biggr]
    -\frac{\sigma^2\ell^2}{2}A_\ell\;.
  \end{split}
\end{equation}
The corresponding Fokker--Plank equation reads
\begin{equation}
  \label{eq:model_fp_pde}
  \begin{split}
    \frac{\partial v}{\partial t}=&-\frac{\partial}{\partial \phi}
    \Biggl\{
    v\sum_{n=1}^2\tilkappa_n \mean{\sin[n(\psi-\phi)]}_\psi \\
    {}&+
    \epsilon v\sin 3\phi\Biggr\}
    +\frac{\sigma^2}{2}\frac{\partial^2v}{\partial\phi^2},
  \end{split}
\end{equation}
where $\mean{\sin[n(\psi-\phi)]}_\psi=
\int_0^{2\pi}v(\psi,t)\sin[n(\psi-\phi)]\mydd\psi$. Accounting this form,
we can reconstruct the respective microscopic equation~\eqref{eq:ens}
for the ensemble of phase oscillators as
\begin{equation}
  \label{eq:model_phs}
  \dot\phi_k=\epsilon\sin(3\phi_k)+
  \sum_{n=1}^2
  \frac{\tilkappa_n}{K}
  \sum_{j=1}^K\sin[n(\phi_j-\phi_k)]+\sigma\xi_k(t).
\end{equation}
In all these equations the
blinking coupling $\tilkappa_n(t)$ is given by expression \eqref{eq:blcoupl}.

The essential control parameters both for the
ensemble~\eqref{eq:model_phs}, and for the Fokker-Plank
system~\eqref{eq:model_fp} are $\kappa_1$ and $\kappa_2$, which have
to satisfy the condition~\eqref{eq:cluster_gr_cond} for clusters to
grow at the corresponding stages.

\subsection{Nonlinear Fokker-Plank equation}

First, we present numerics for the Fokker-Plank
system~\eqref{eq:model_fp}. Since this system appears as a spectral
decomposition of the nonlinear integro-differential
equation~\eqref{eq:model_fp_pde}, it is stiff, and we integrate it
numerically using the method of exponential time differencing, see
Ref.~\cite{Cox2002430}. As follows from Eq.~\eqref{eq:order_prm},
$|A_\ell|\leq 1$. Due to the terms proportional to $\ell^{-2}$, the
amplitudes of $A_\ell$ decay fast with $\ell$. Direct verification
shows that already for $\ell=5$ the magnitude is small enough,
$|A_5|\ll1$.  Thus we cut the infinite set of Eqs.~\eqref{eq:model_fp}
at $\ell_\inxcut=10$.

\begin{figure*}
  \widfig{1}{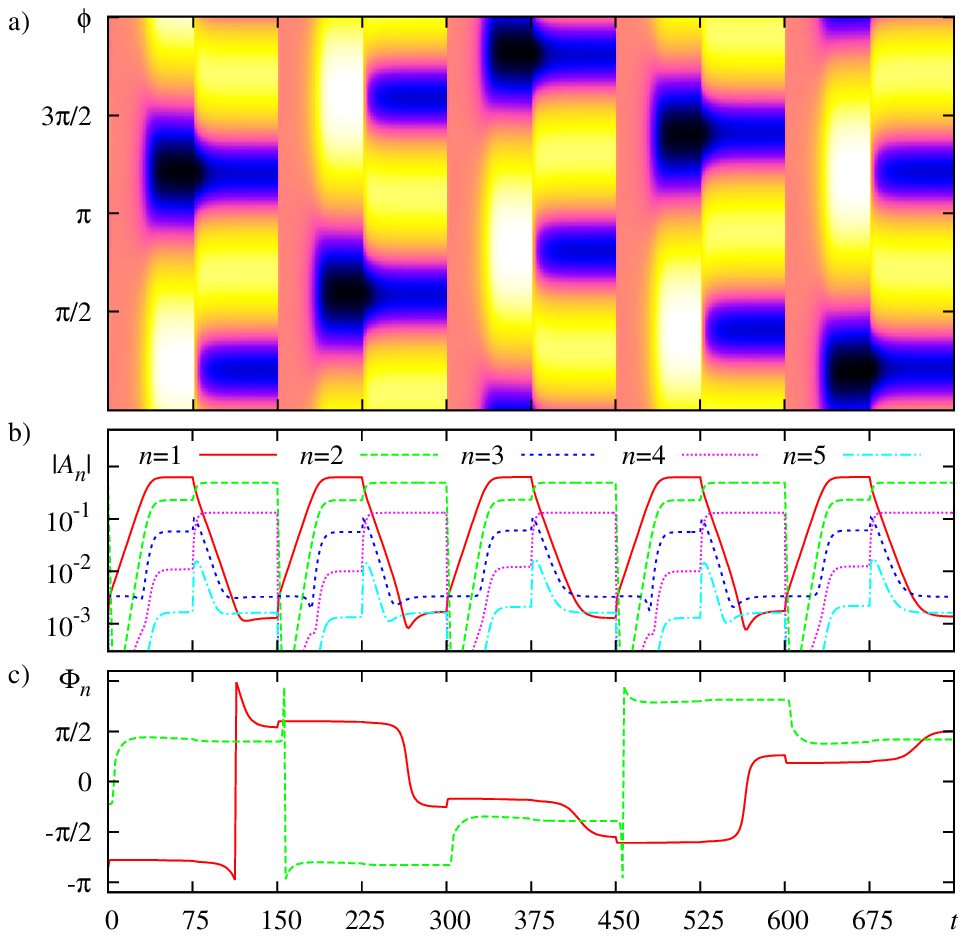}
  \widfig{1}{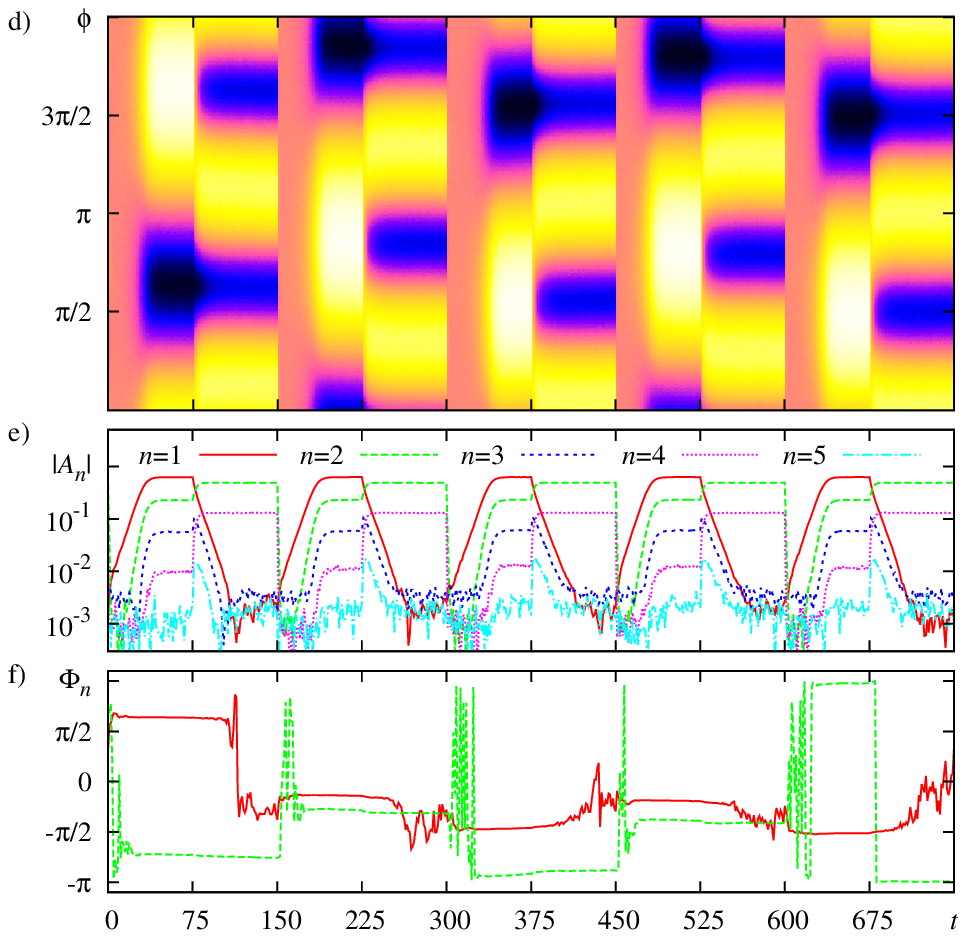}
  \caption{\label{fig:sptm_fp_phs} (color online) a) Probability density
    $v(\phi,t)$ determined by~\eqref{eq:prob_dens}, where $A_n(t)$ are
    computed according to Eq.~\eqref{eq:model_fp}; d) distribution for
    the ensemble of $\phi_k$, see Eq.~\eqref{eq:model_phs}, where the
    total number of the oscillators is $K=10^6$. Darker colors
    correspond to higher values of $v(\phi,t)$. Five periods of
    switching of $\tilkappa_{1,2}(t)$ are shown. b) and c) Absolute
    values and arguments of $A_n$ computed according to
    Eq.~\eqref{eq:model_fp}; e) and f) the same computed via
    Eq.~\eqref{eq:order_prm} for the ensemble. $|A_n|$ are plotted in
    logarithmic scale. Parameters: $\kappa_1=1.3$, $\kappa_2=2.3$,
    $T=150$, $\epsilon=0.01$, $\sigma=1$, $\ell_\inxcut=10$.}
\end{figure*}

Figure~\ref{fig:sptm_fp_phs}(a) illustrates the temporal behavior of the
probability density $v(\phi,t)$ computed according to
Eq.~\eqref{eq:prob_dens}. At the stage 1 ($0\leq t\leq 75$) the
$1$-cluster develops.  Then, after the switch to the stage 2 at
$t=T/2=75$, the density acquires two humps. One part of the
$2$-cluster is located just at the site of the former $1$-cluster, and
the second one appears shifted by $\pi$. As we already discussed, this
situation corresponds to the quadratic relation for the complex
amplitudes $A_2\sim A_1^2$ and to the doubling of the argument. At the
initial epoch of the next stage 1 at $t\gtrsim 150$, the $2$-cluster
disappears, and the $1$-cluster reappears.  Notice that the hump
emerges now at a site which is different both from that of the
previous $2$-cluster, and that of the former $1$-cluster.  It is so
because of the presence of a small force $f(\phi)$ corresponding to
the term proportional to $\epsilon$ in Eqs.~\eqref{eq:model_fp}.

Figures~\ref{fig:sptm_fp_phs}(b) and (c) show magnitudes and arguments of
$A_n$. At stage 1 all the magnitudes $|A_n|$ grow and saturate at some
level; in this state the main mode $A_1$ dominates. At stage 2 the
mode $A_2$ grows, together with its harmonics, while all odd modes
decay.

Let us now focus on the phase dynamics. The phase $\argA_2$ does not
change at the transition from stage 1 to stage 2 at $t=75$.  At the
beginning of stage 2 the phase of the first mode $\argA_1$ is nearly
constant while the amplitude of this mode remains relatively large;
however at $t\approx 100$ the amplitude drops to level $\sim\epsilon$,
and now this mode becomes to be driven by $\epsilon A_2^*$. Here, the
mode $A_1$ accepts the phase $-\argA_2$ that can be seen in
Fig.~\ref{fig:sptm_fp_phs}(c) as a sharp transition around $t\approx
110$. After this event, the phase of the mode $A_1$ does not vary
significantly. At $t=T=150$ the new stage 1 starts, here the amplitude
of the previously active mode $A_2$ decays rapidly, but it reappears
soon as the second harmonics of the mode 1, accepting its doubled
phase $2\argA_1$ (see the jump of $\argA_2$ at $t\approx 155$ followed
by the growth of $|A_2|$).

\begin{figure}
  \widfig{1}{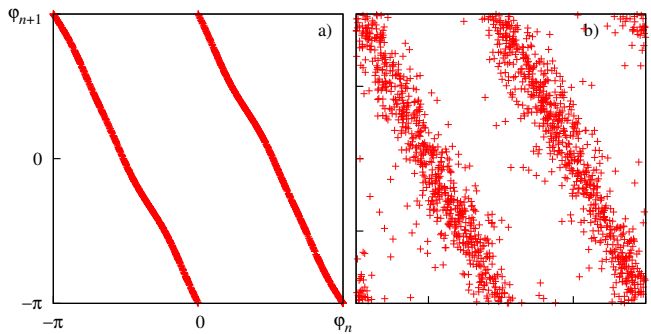}
  \caption{\label{fig:fmap}a) The plot $\varphi_{n+1}$ vs. $\varphi_n$
    where $\varphi_n=\arg A_1(nT+\frac{9}{20}T)$, and $A_\ell$ is a
    solution of Eqs.~\eqref{eq:model_fp}. b) The same diagram for the
    ensemble~\eqref{eq:model_phs}, where $A_1$ is computed according
    to Eq.~\eqref{eq:order_prm}. Parameters correspond to
    Fig.~\ref{fig:sptm_fp_phs}.}
\end{figure}

Although the dynamics of the phases and the amplitudes observed in
Fig.~\ref{fig:sptm_fp_phs}(a,b,c) nicely corresponds to the qualitative
picture of section~\ref{sec:esd}, it is important to verify that the
phase evolution really follows the Bernoulli-like map~\eqref{eq:ber}.
In Fig.~\ref{fig:fmap}(a) a diagram is shown for the values of
argument of the complex variable $A_1$ recorded stroboscopically at
$t=t_n=(n+\frac{9}{20})T$. Indeed, the plot demonstrates the doubling
of the phase, as expected.
Thus, actually, the phases of the alternately
arising clusters evolve according to
the expanding circle map that implies the hyperbolic chaos.

\subsection{Ensemble of noisy oscillators}

Here we report on the results of a straightforward simulation of the
ensemble dynamics according to Eqs.~\eqref{eq:model_phs}.  We perform
numerical solutions for these stochastic differential equations
employing a special version of the Runge-Kutta-type algorithm
suggested in~\cite{Honeycutt92}.  Figure~\ref{fig:sptm_fp_phs}(d)
illustrates the temporal behavior of the probability density
$v(\phi,t)$ computed from the instantaneous distributions of the
phases $\phi_k$ for the elements of the ensemble, and the panels (e)
and (f) demonstrate the dynamics of the mean-field variables
$A_n(t)$. The parameters are the same as in the previous simulations
of the Fokker-Plank equation. Except for the obvious difference in the
initial conditions, the distributions $v(\phi,t)$ and the magnitudes
of $A_n(t)$ at their high levels behave visually similar. The dynamics
of the phases is definitely more noisy than that in
Fig.~\ref{fig:sptm_fp_phs}(c), especially in the regions where the
magnitudes of the corresponding modes are small: in these regions even
small finite-size fluctuations produce a large effect on the phases.
When the magnitude of $A_1(t)$ is high (see for example the time
interval $30<t<75$), its argument does not respond to the noise. But
when $A_1$ gets small, the argument fluctuates (see the time interval
$110<t<150$). The same is true for $A_2$. The most harmful are the
fluctuations of $\argA_1$ because this variable inherits the argument
value from $A_2$ when $|A_1|$ is small. If the fluctuations are strong
enough, they wash out the correct value of the argument, and the phase
multiplication mechanism ceases to operate properly. The fluctuations
of $A_2$ are not so essential since they occur only in epochs, when
the component $A_2$ does not play an important role. One can see from
the figure that the phase transfer happens when $A_2$ has a large
magnitude and thus is not affected by the noise (see the time interval
$180<t<225$).

Figure~\ref{fig:fmap}(b) shows the diagram for the stroboscopically
recorded values of $\argA_1$ corresponding to the dynamical behavior
illustrated in Fig.~\ref{fig:sptm_fp_phs}(d,e,f). One can observe that the
phases demonstrate the expected Bernoulli-like doubling mapping in
average, though the fluctuations produce a noticeable statistical
widening (which decreases as the size of the ensemble $K$ grows).

\subsection{Lyapunov exponent and hyperbolicity test}

Now we return to the Fokker--Plank system~\eqref{eq:model_fp} and
discuss how the Lyapunov exponents (LEs) depend on $\kappa_1$ and
$\kappa_2$. Also we test hyperbolicity of the chaotic dynamics of
the system~\eqref{eq:model_fp}.

\begin{figure}
  \widfig{1}{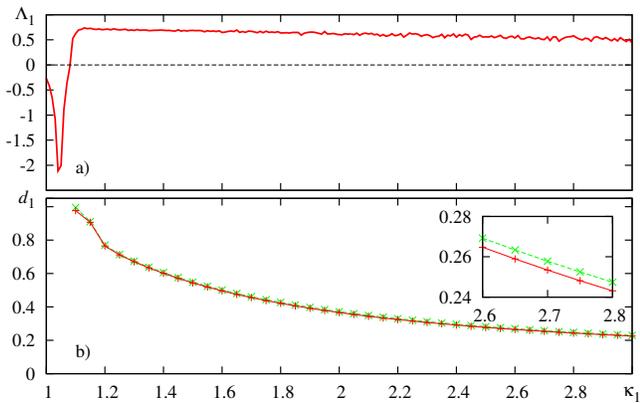}
  \caption{\label{fig:lydet_kap1}(a) The first Lyapunov exponent, and
    (b) the lower and upper boundaries of the distribution of $d_1$
    vs. $\kappa_1$. The data relate to the stroboscopic map
    corresponding to the Fokker-Plank system~\eqref{eq:model_fp} at
    $t_n=(n+\frac{9}{20})T$ . Distributions of $d_1$ for each
    $\kappa_1$ were computed for $10^4$ attractor
    points. $\kappa_2=2.3$, other parameters correspond to
    Fig.~\ref{fig:sptm_fp_phs}.}
\end{figure}

We calculate the Lyapunov exponents in the deterministic system~\eqref{eq:model_fp} with
the standard Benettin algorithm. Because the system is periodically driven, we report
as LEs the values $\Lambda=\lambda T$, i.e. the ``Lyapunov multipliers'' over the period.
In all cases we have found at most one positive LE, the others were negative.
Figure~\ref{fig:lydet_kap1}(a) shows the largest LE versus  parameter $\kappa_1$;
in a large range of this parameter it is positive and
 close to the expected value $\ln 2\approx
0.69314$. The other exponents are all negative, and their magnitudes are
much higher. For example, at $\kappa_1=1.3$, and $\kappa_2=2.3$ we have
$\Lambda_i=\{0.688, -17.355, -197.507, -245.781, -705.748\}$, and the
corresponding Kaplan-Yorke dimension is $1.04$.

\begin{figure}
  \widfig{1}{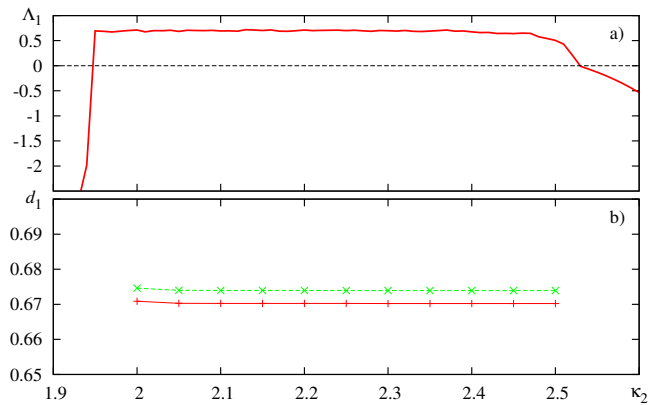}
  \caption{\label{fig:lydet_kap2} The same quantities as in
    Fig.~\ref{fig:lydet_kap1} against $\kappa_2$, while
    $\kappa_1=1.3$.}
\end{figure}

The positive Lyapunov exponent varies slowly depending on the
parameter, but no tips or dips, like in many non-hyperbolic systems,
are observed.  This may be regarded as a manifestation of the
structural stability intrinsic to the hyperbolic dynamics.
Figure~\ref{fig:lydet_kap2}(a) represents dependence of $\Lambda$ on
$\kappa_2$, while $\kappa_1$ is constant. Again, the same features are
observed: the largest Lyapunov exponent is close to the value $\ln 2$,
and it does not demonstrate any notable variation within a wide range
of the parameter.

To test hyperbolicity, we employed the method which has been specially
developed for high-dimensional systems, see Ref.~\cite{FastHyp12} for
details and Ref.~\cite{CLV2012} for the mathematical background. In
brief, the method includes the computation of the first $k$
orthogonalized Lyapunov vectors moving forward in time (so called
Gram-Schmidt vectors or backward Lyapunov vectors~\cite{CLV2012}), and
also the first $k$ orthogonalized Lyapunov vectors obtained moving
backward in time from the conjugate variational equations (forward
Lyapunov vectors); here $k$ is the dimension of the unstable
manifold. Given these vectors, a $k\times k$ matrix of their scalar
products is built. Its smallest singular value $d_k$ is cosine of the
angle between the expanding tangent subspace and the orthogonal
complement to the contracting tangent subspace, and thus $d_k$ can be
an indicator of hyperbolicity. For a non-hyperbolic case, the $d_k$
values vanish somewhere along the trajectories (as the tangencies
between the associated expanding and contracting subspaces occur), but
for situations of the hyperbolic dynamics the distribution of $d_k$ is
well separated from zero.

In Fig.~\ref{fig:lydet_kap1}(b) we illustrate the results of this
hyperbolicity test. In our case the expanding subspace is
one-dimensional.  Thus, for each $\kappa_1$ we compute $10^4$ points
on the attractor of the stroboscopic map, and evaluate the indicator
$d_1$ at each of them. Then, the smallest and the largest values of
$d_1$ are plotted on the diagram, marked with crosses and pluses,
respectively. The first observation is that the smallest $d_1$ is very
far from zero; this confirms the hyperbolicity. Another noteworthy
observation is that the maximal and the minimal values of $d_1$ are
very close to each other. This means that the mutual orientations of
the expanding and contracting directions remain practically unaltered
on the entire attractor. This can be treated as another manifestation
of the structural stability. Figure~\ref{fig:lydet_kap2}(b) shows the
hyperbolicity indicator $d_1$ in dependence on $\kappa_2$, it is again
well separated from zero.

\subsection{Lyapunov exponents for the micro-level dynamics}

It is instructive to compare the macroscopic LEs for the dynamics of
the order parameters with microscopic LEs describing stability of the
dynamics of individual oscillators.  A direct approach to computation
of the full spectrum of Lyapunov exponents for the ensemble of phase
oscillators~\eqref{eq:model_phs} can easily exhaust a computer memory
and take an extremely long time. However, there is a simple way to
estimate microscopic LEs assuming their decoupling from the
macroscopic dynamics.  Neglecting small $\epsilon$ in
Eq.~\eqref{eq:order_prm}, we can rewrite this equation as
\begin{equation}
  \label{eq:model_phs_fp1}
  \dot\phi_k=\frac{1}{2\myii}
  \sum_{n=1}^2
  \tilkappa_n (A_ne^{-\myii n\phi_k}-A_n^*e^{\myii n\phi_k})+\sigma\xi_k(t)\;.
\end{equation}
Let us now assume that on stages 1 and 2 the order parameters $A_1$ and $A_2$ are roughly constant.
Then the variations of the phases satisfy
\begin{equation}
  \label{eq:model_phs_fp1-lin}
  \frac{1}{\delta\phi_k}\frac{d}{dt}\delta\phi_k=-\sum_{n=1}^2 n\tilkappa_n(A_ne^{-\myii n\phi_k}+A_n^*e^{\myii n\phi_k}).
\end{equation}
Now the microscopic LE is evaluated as an average of the r.h.s. of Eq.~\eqref{eq:model_phs_fp1-lin} multiplied by $T/2$, which
yields $\Lambda_{\text{mic}}=-\frac{T}{2}\sum_{n=1}^2 n\tilkappa_n |A_n|^2$. Substituting here the approximate
expressions for stationary magnitudes of the order parameters, valid for small subcritical couplings $\kappa_1\gtrsim
\sigma^2$, $\kappa_2\gtrsim 2\sigma^2$, we obtain the following estimate:
\begin{equation}
  \label{eq:model_phs_lyap_estim}
  \Lambda_{\text{mic}}\approx -T\sigma^2\left(\frac{\kappa_1-\sigma^2}{\kappa_1}+4\frac{\kappa_2-2\sigma^2}{\kappa_2}\right)\;.
\end{equation}
For the parameters used above, $\sigma=1,\kappa_1=1.3,\kappa_2=2.3,T=150$, this yields $ \Lambda_{\text{mic}}\approx-113$ in a reasonable agreement with the numerics.

Thus, contrary to many situations where chaos occurs in the
micro-level description, while the macro-parameters manifest rather
regular behavior, here we observe an impressively opposite case: the
micro-dynamics is stable (because $ \Lambda_{\text{mic}}<0$), but at
the macro-level the hyperbolic chaos takes place.

\section{Conclusion}
In this paper we have demonstrated that blinking coupling in populations
of noisy oscillators may result in hyperbolic chaos of macroscopic mean fields.
The blinking is between two simplest modes of coupling: one leads to a one-cluster synchronized state, another one leads to appearance of two clusters, shifted by $\pi$. Correspondingly, at these
couplings two different order parameters dominate: the usual Kuramoto order parameter for the
one-cluster state, and the second-harmonics Daido order parameter for the two-cluster state.
While the magnitudes of these order parameters show no significant variations from cycle to cycle, the phases obey a strongly chaotic Bernoulli transformation, and thus demonstrate
a hyperbolic chaos. Noteworthy, there is a realistic situation where the two types of global coupling
(one-cluster and two-cluster) can be observed: this is an ensemble of pendula hanging on a common beam.
In such a configuration with two pendulum clocks Ch. Hyugens observed synchronization  more than 300 years ago~(see a translation of his notes in~\cite{Pikovsky-Rosenblum-Kurths-01}). Nowadays one reproduces these experiments with metronomes~\cite{Kapitaniak_etal-12}. The horizontal motions of the beam result in the
one-cluster coupling, while
the vertical motions lead to the two-cluster coupling~\cite{Czolczynski_etal-13}.

Our analysis is mostly based on the consideration of the thermodynamic limit, where
the population can be described by the nonlinear Fokker-Planck equation.
We verify hyperbolic chaos in this integral-differential equation, by direct simulations of equations for
the modes and showing that their phases obey a Bernoulli map, and by checking that the stable and the
unstable directions in the tangent space are
never  tangent to each other. For the original population a direct simulation gives a picture very similar
to that for the nonlinear Fokker-Planck equation, but the one-dimensional transformation of the phases looks like a noisy Bernoulli map, due to finite-size
effects.

In a more general perspective, the nonlinear Fokker-Planck equation,
as an  integro-differential equation with partial derivatives, is a
representative of a class of deterministic distributed systems demonstrating
pattern formation. Different cluster synchronization states correspond to
different patterns; phases of clusters correspond to the spatial
positions of the patterns. With this interpretation, our results demonstrate that
an interaction between blinking patterns results in a chaotic relocation of their
positions, moreover, this chaos is hyperbolic. A kind of such behavior was
reported recently in a model of interacting Turing
patterns with different wave numbers~\cite{HypTur2012}.
The presented model based on the nonlinear
Fokker--Plank equation provides another indication that the above
mechanism of the hyperbolic chaos is realizable in quite generic circumstances.

The authors acknowledge support from the RFBR grant No 11-02-91334 and
DFG grant No PI 220/14-1.

\bibliography{hans}

\end{document}